\documentclass[12pt]{article}
\usepackage{amsmath, amssymb, graphicx, float}
\usepackage{geometry}
\usepackage{hyperref}
\usepackage{cite}  
\usepackage{enumitem}    % For customizing lists
\usepackage{geometry}    % Optional: Adjust page margins
\usepackage{hyperref}    % Optional: For hyperlinks
\usepackage{graphicx}    % For including graphics
\usepackage{caption}
\usepackage{subcaption}
\graphicspath{{images/}} % If your images are in the 'images' directory
\title{Magnetic Field Data Calibration with Transformer Model Using Physical Constraints: A Scalable Method for Satellite Missions, Illustrated by Tianwen-1}
\author{Beibei Li, Yutian Chi, Yuming Wang}
\date{}

\begin{document}

\maketitle

\begin{abstract}

%This study introduces a novel approach that integrates the magnetic field data correction from the Tianwen-1 Mars mission with a neural network architecture constrained by physical principles derived from Maxwell’s equations. By employing a Transformer-based model capable of efficiently handling sequential data, the method corrects measurement anomalies caused by satellite dynamics, instrument interference, and environmental noise. As a result, it significantly improves both the accuracy and the physical consistency of the calibrated data. Compared to traditional methods that require long data segments and manual intervention often taking weeks or even months to complete this new approach can finish calibration in just minutes to hours, and predictions are made within seconds. This innovation not only accelerates the process of space weather modeling and planetary magnetospheric studies but also provides a robust framework for future planetary exploration and solar wind interaction research.

This study introduces a novel approach that integrates the magnetic field data correction from the Tianwen-1 Mars mission with a neural network architecture constrained by physical principles derived from Maxwell’s equations. By employing a Transformer based model capable of efficiently handling sequential data, the method corrects measurement anomalies caused by satellite dynamics, instrument interference, and environmental noise. In addition to the original approach, we implement two distinct Transformer architectures: a standard Transformer and a physics informed Transformer. The latter incorporates a Fourier Transform branch to extract frequency domain features and enforces a divergence free constraint on the predicted magnetic field, ensuring physical consistency. As a result, while both models significantly improve accuracy and reduce calibration time from weeks or months to minutes or hours. This innovation not only accelerates the process of space weather modeling and planetary magnetospheric studies but also provides a robust framework for future planetary exploration and solar wind interaction research.
\end{abstract}

\section{Introduction}
Magnetic field measurements from space missions, such as Tianwen-1, are critical for exploring magnetospheric interactions and solar wind dynamics. However, magnetometer data often suffer from disturbances caused by satellite dynamics, onboard instrument interference, and environmental noise. For instance, changes in satellite orientation can lead to anomalies in magnetic field measurements due to interference from electric currents within the satellite's instruments. These disturbances necessitate careful data correction to ensure the accuracy and reliability of measurements.

Traditional correction methods rely heavily on human expertise and are rooted in well established physical and mathematical principles. While these methods have proven effective, they are inherently limited by their long processing times and delays in real time prediction\cite{zou2023inflight} \cite{wang2024calibration} \cite{olsen2020magnetic} \cite{gebre-egziabher2002calibration} \cite{farrell1995method}. In contrast, machine learning models, though rarely applied in this field, offer strong predictive capabilities and the potential for faster computations. This study seeks to address these limitations by combining the strengths of traditional correction methods with the adaptability and efficiency of machine learning models, thereby improving timeliness while ensuring both physical consistency and improved real time performance. This study bridges the gap between data driven models and physics based understanding by integrating Maxwell's equations into the neural network architecture as physical information. The key innovations are:
\begin{enumerate}
    \item \textbf{Magnetic Field Data Correction:} A systematic calibration pipeline that accounts for instrument drift, orientation changes, and environmental noise.
    \item \textbf{Neural Network Integration with Physical Information:} The integration of physical fields, such as electric and magnetic fields, along with real time equation calculations, directly into the neural network as input features.  We introduce two Transformer based architectures. The first is a standard Transformer model for time series prediction, and the second is a physics informed Transformer that integrates a Fourier Transform branch and enforces a divergence free constraint via a dedicated physics layer. These additional components enable the model to capture both time domain and frequency domain features, while simultaneously integrating Maxwell’s equations into the learning process. This dual model strategy bridges the gap between data driven methods and physics based understanding.
    \item \textbf{Significant Improvement in Computational Efficiency:} The proposed neural network approach significantly improves computational efficiency, reducing the calibration and correction process from several days or months to just a few minutes or hours. 
    \item \textbf{Scalability:} This method demonstrates strong scalability and can be effectively adapted for the correction of magnetic and electric field interference, as well as signal distortions, across various satellite missions and measurement systems.
\end{enumerate}

\section{Methods}

\subsection{Traditional Magnetometer Calibration Methods}
Traditional magnetometer calibration methods rely heavily on physical models and predefined mathematical approaches. These methods correct for instrumental offsets, dynamic fields generated by spacecraft systems, and environmental interferences in magnetic measurements \cite{zou2023inflight} \cite{olsen2020magnetic} \cite{ness1971use} \cite{gebre-egziabher2002calibration} \cite{farrell1995method}.
\begin{itemize}
    \item \textbf{Instrument Drift Correction:} Instrumental drift occurs due to aging, thermal variations, and operational wear, gradually altering the offset values of magnetometer readings. Correction is achieved by comparing in flight measurements with known reference conditions, often requiring extensive post mission data analysis.
    
    \item \textbf{Dynamic and Static Field Correction:} Dynamic fields generated by onboard spacecraft systems are identified and removed using dual sensor methods \cite{ness1971use}, separating spacecraft induced fields from natural magnetic measurements. Static fields and instrumental offsets are treated together as "zero offset" values, which require periodic calibration during flight.
    
    \item \textbf{Environmental Noise Filtering:} External noise from solar wind and other sources is minimized through filtering techniques, such as moving averages and variance analysis, to maintain measurement accuracy.
    
    \item \textbf{Physical Phenomenon Based Methods:}
    \begin{itemize}
        \item \textbf{Alfvén Wave Analysis:} Leveraging natural periodicities in interplanetary magnetic fields to determine offsets \cite{wang2024calibration}.
        \item \textbf{Mirror Mode Structures and Current Sheets:} Specific magnetic structures are used as calibration references, but they are less frequent in planetary environments like Mars compared to the solar wind \cite{wang2024calibration}.
    \end{itemize}
\end{itemize}

\textbf{Key Challenges}
\begin{itemize}
    \item \textbf{Extended Time Requirements:} Traditional methods can take weeks or months for complete calibration due to the need for long data segments and manual intervention.
    
    \item \textbf{Dependence on Specific Conditions:} Calibration often depends on favorable conditions, such as solar wind periods, limiting applicability in planetary magnetosheaths or other complex regions.
    
    \item \textbf{Segment Specific Analysis:} Data gaps and variability require separate calibration for each continuous data segment, increasing computational and operational overhead.
\end{itemize}

\subsection{Machine Learning Methods}

\subsubsection{Data}

Magnetometer data are affected by external factors such as satellite orientation changes and onboard instrument interference. Data from both inner and outer magnetometer probes, as well as their differences, are utilized for correction.
Synchronization of data timestamps from multiple sensors, identification and labeling of anomalous data, and resampling at one minute intervals to reduce noise and computational complexity.
We have all the original data from Tianwen-1, including the corrected and published datasets at \url{https://space.ustc.edu.cn/dreams/tw1_momag/?magdata=cal&sr=1}.

\subsubsection{Input}
The input data for the neural network model comprises the following components:
\begin{itemize}
    \item \textbf{Internal Probes:}
    Three columns representing the magnetic field measurements obtained from the internal probes.
    \item \textbf{External Probes:}
    Three columns representing the magnetic field measurements obtained from the external probes.
    
    \item \textbf{Differences:}
    Three columns representing the differences between the internal and external magnetic field measurements.
    
    \item \textbf{Satellite Position:}
    Three columns detailing the satellite's position coordinates.
    
    \item \textbf{Satellite Attitude:}
    Three columns detailing the satellite's attitude parameters.
    
    \item\textbf{Calculation of Electric Field Components:} Magnetic field data (\(B_x\), \(B_y\), \(B_z\)) are used to calculate electric field components (\(E_x\), \(E_y\), \(E_z\)) using Maxwell's equations
    \[
\frac{\partial B_z}{\partial y} - \frac{\partial B_y}{\partial z},\  \frac{\partial B_x}{\partial z} - \frac{\partial B_z}{\partial x},\  \frac{\partial B_y}{\partial x} - \frac{\partial B_x}{\partial y}
\]
.
%    \[
%    \nabla \times \mathbf{E} = -\frac{\partial \mathbf{B}}{\partial t}.
%    \]

%To calculate the electric field components based on the magnetic field, we apply Faraday's law of induction. In the case of spatial variations, the electric field components can be expressed as:

%\[
%\begin{aligned}
%E_x &= - \left( \frac{\partial B_z}{\partial y} - \frac{\partial B_y}{\partial z} \right), \\
%E_y &= - \left( \frac{\partial B_x}{\partial z} - \frac{\partial B_z}{\partial x} \right), \\
%E_z &= - \left( \frac{\partial B_y}{\partial x} - \frac{\partial B_x}{\partial y} \right).
%\end{aligned}
%\]

The Ampère-Maxwell Law is a fundamental equation in Maxwell's equations that describes the relationship between the magnetic field \( \mathbf{B} \), the current density \( \mathbf{J} \), and the time rate of change of the electric field \( \frac{\partial \mathbf{E}}{\partial t} \). Understanding this relationship is crucial for analyzing time varying electromagnetic fields and the propagation of electromagnetic waves.

The differential form of the Ampère-Maxwell Law is given by:

\begin{equation}
\nabla \times \mathbf{B} = \mu_0 \mathbf{J} + \mu_0 \varepsilon_0 \frac{\partial \mathbf{E}}{\partial t}
\label{ampere_maxwell}
\end{equation}

where:
\begin{itemize}
    \item \( \nabla \times \mathbf{B} \) is the curl of the magnetic field,
    \item \( \mu_0 \) is the permeability of free space,
    \item \( \mathbf{J} \) is the current density,
    \item \( \varepsilon_0 \) is the permittivity of free space,
    \item \( \frac{\partial \mathbf{E}}{\partial t} \) is the time rate of change of the electric field.
\end{itemize}

To derive an expression for the electric field \( \mathbf{E} \) from the Ampère-Maxwell Law, follow these steps:

Starting with Equation \eqref{ampere_maxwell}, we can solve for the time derivative of the electric field:

\begin{equation}
\mu_0 \varepsilon_0 \frac{\partial \mathbf{E}}{\partial t} = \nabla \times \mathbf{B} - \mu_0 \mathbf{J}
\end{equation}

Dividing both sides by \( \mu_0 \varepsilon_0 \):

\begin{equation}
\frac{\partial \mathbf{E}}{\partial t} = \frac{1}{\mu_0 \varepsilon_0} \left( \nabla \times \mathbf{B} - \mu_0 \mathbf{J} \right)
\label{electric_field_derivative}
\end{equation}

To obtain the electric field \( \mathbf{E} \) itself, integrate Equation \eqref{electric_field_derivative} with respect to time:

\begin{equation}
\mathbf{E} = \mathbf{E}_0 + \frac{1}{\mu_0 \varepsilon_0} \int \left( \nabla \times \mathbf{B} - \mu_0 \mathbf{J} \right) dt
\label{electric_field}
\end{equation}

where \( \mathbf{E}_0 \) represents the initial electric field.

In a Cartesian coordinate system, the vector equation can be decomposed into its scalar components along the \( x \), \( y \), and \( z \) axes.

Assume the magnetic field \( \mathbf{B} \) and current density \( \mathbf{J} \) have components:

\[
\mathbf{B} = (B_x, B_y, B_z)
\]
\[
\mathbf{J} = (J_x, J_y, J_z)
\]

The curl of \( \mathbf{B} \) is:

\[
\nabla \times \mathbf{B} =
\left( \frac{\partial B_z}{\partial y} - \frac{\partial B_y}{\partial z},\ 
\frac{\partial B_x}{\partial z} - \frac{\partial B_z}{\partial x},\ 
\frac{\partial B_y}{\partial x} - \frac{\partial B_x}{\partial y} \right)
\]

Substituting the curl into Equation \eqref{electric_field_derivative}, we obtain the scalar components:

\[
\frac{\partial E_x}{\partial t} = \frac{1}{\mu_0 \varepsilon_0} \left( \frac{\partial B_z}{\partial y} - \frac{\partial B_y}{\partial z} - \mu_0 J_x \right)
\]

\[
\frac{\partial E_y}{\partial t} = \frac{1}{\mu_0 \varepsilon_0} \left( \frac{\partial B_x}{\partial z} - \frac{\partial B_z}{\partial x} - \mu_0 J_y \right)
\]

\[
\frac{\partial E_z}{\partial t} = \frac{1}{\mu_0 \varepsilon_0} \left( \frac{\partial B_y}{\partial x} - \frac{\partial B_x}{\partial y} - \mu_0 J_z \right)
\]

Integrate each component with respect to time to obtain the expressions for the electric field components:

\[
E_x = E_{x0} + \frac{1}{\mu_0 \varepsilon_0} \int \left( \frac{\partial B_z}{\partial y} - \frac{\partial B_y}{\partial z} - \mu_0 J_x \right) dt
\]

\[
E_y = E_{y0} + \frac{1}{\mu_0 \varepsilon_0} \int \left( \frac{\partial B_x}{\partial z} - \frac{\partial B_z}{\partial x} - \mu_0 J_y \right) dt
\]

\[
E_z = E_{z0} + \frac{1}{\mu_0 \varepsilon_0} \int \left( \frac{\partial B_y}{\partial x} - \frac{\partial B_x}{\partial y} - \mu_0 J_z \right) dt
\]

where \( E_{x0} \), \( E_{y0} \), and \( E_{z0} \) are the initial components of the electric field.

We understand the relationship between the magnetic and electric fields, and we have measurements of $B_x$, $B_y$, and $B_z$. Changes in the satellite's orientation can induce variations in the electric field, which contribute to anomalies in our data. Therefore, we incorporate this known information—specifically, the electric field derived from magnetic field measurements—as inputs to our neural network model
 \[
\frac{\partial B_z}{\partial y} - \frac{\partial B_y}{\partial z},\  \frac{\partial B_x}{\partial z} - \frac{\partial B_z}{\partial x},\  \frac{\partial B_y}{\partial x} - \frac{\partial B_x}{\partial y}
\].
\end{itemize}
We try different combination of inputs, the current inputs give us the best results. 

\subsubsection{Output}
The output data consists of the corrected magnetic field measurements. This data is expected to refine the original measurements by mitigating noise and compensating for any systematic errors, thereby providing more accurate representations of the magnetic fields.
With the introduction of two transformer models, the standard Transformer outputs magnetic field values, while the physics informed Transformer further refines these predictions by integrating electric field computations and enforcing the divergence free condition. 
\subsubsection{Neural Network Architecture}
\begin{figure}[H]
  \centering
  \includegraphics[width=\textwidth]{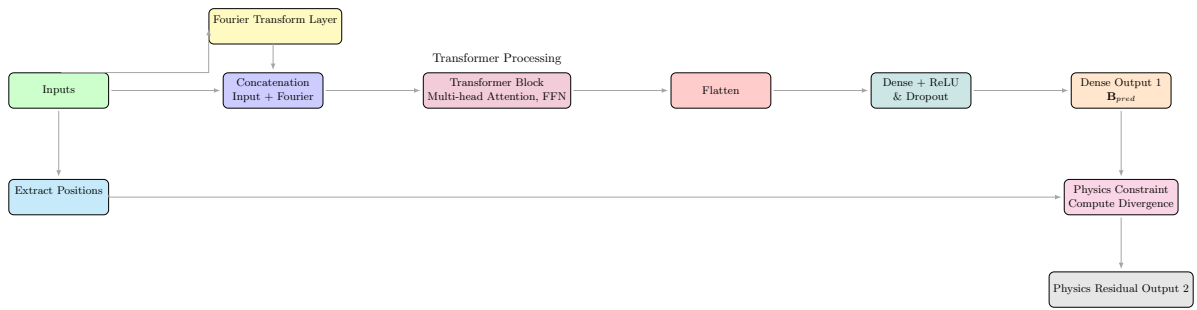}
  \caption{Physics Informed Transformer with Fourier Transform and Physics Constraint}
  \label{fig:Transformer}
\end{figure}

\paragraph{Transformer}
The neural network model is a Transformer\cite{vaswani2017attention} for time series prediction. It features a custom \texttt{TransformerBlock} that utilizes multi-head attention to capture dependencies across time steps. This block includes a feedforward network with ReLU activation, layer normalization, dropout regularization, and residual connections to improve training stability and model capacity. The complete architecture comprises an input layer, the TransformerBlock, a flattening layer, a dense layer with ReLU activation, dropout, and an output layer that predicts three target variables. Data preprocessing involves resampling the time series, engineering features by computing derivatives and electric fields, standardizing features and targets, and constructing input sequences. The model is trained using the Adam optimizer with early stopping based on validation loss.\\
\paragraph{Physics Informed Transformer with Fourier Transform and Physics Constraint}
The second architecture in Figure\ref{fig:Transformer} extends the basic Transformer design by incorporating both a Fourier transformation branch and a physics informed constraint. Specifically, a Fourier transformation layer is applied to the entire input sequence to extract frequency domain features, which are then concatenated with the original input data. This augmented input is processed through a custom \texttt{TransformerBlock} that leverages multi-head attention and a feedforward network with dropout and layer normalization. After flattening and passing through additional dense layers, the model outputs predictions for the magnetic field components. Moreover, position features are extracted from the input and used in a \texttt{PhysicsConstraintLayer} to compute the divergence of the predicted magnetic field, enforcing the physical principle of zero divergence in magnetohydrodynamics. The combined loss function during training optimizes both the prediction accuracy and the adherence to the physical constraint.
\subparagraph{Fourier Transform Layer Structure}
The Fourier Transform layer is designed to extract frequency domain features from the input time series data. The process involves the following steps: The input data is first converted to a complex format to enable complex valued operations. A Fast Fourier Transform is applied to transform the data into its frequency domain representation.  The real and the imaginary part of the FFT result is extracted and used as the output of the layer. 

%\begin{enumerate}
%    \item The input data is first converted to a complex format to enable complex-valued operations.
%    \item A Fast Fourier Transform (FFT) is applied to transform the data into its frequency domain representation.
%    \item An Inverse Fast Fourier Transform (IFFT) is then performed to convert the data back to the time domain.
%    \item The real part of the IFFT result is extracted and used as the output of the layer.
%\end{enumerate}
%This design allows the model to incorporate frequency-domain insights, which can help capture periodic or oscillatory patterns within the time series data.

\subparagraph{Physics Informed Structure}
The physics informed structure integrates prior physical knowledge by enforcing a divergence free constraint on the predicted magnetic field. This is achieved through a dedicated layer that computes the divergence residual to ensure that the predicted magnetic field, \(\mathbf{B}_{pred}\), satisfies \(\nabla \cdot \mathbf{B} = 0\). The procedure is as follows:
\begin{enumerate}
    \item Extract the positional features from the input data and combine them with the predicted magnetic field values.
    \item Compute the Jacobian matrix of the predicted magnetic field with respect to the spatial coordinates.
    \item Calculate the divergence by taking the trace of the Jacobian matrix.
    \item Include this divergence as a residual in the loss function to penalize deviations from the physical law.
\end{enumerate}
This physics informed constraint not only improves the model's accuracy in predicting the magnetic field but also ensures that the predictions adhere to fundamental physical principles.
\section{Results}
The calibration pipeline not only effectively reduced noise and anomalies but also drastically decreased data correction time from several days or even months to just tens of minutes or hours. Magnetic field data corrections showed strong agreement with theoretical models and observational benchmarks.

The Transformer model achieved a mean absolute error 0.513 nT. By incorporating electric and magnetic field components as inputs, the model demonstrated improved performance and physical consistency.

\begin{figure}[H]
  \centering
  \includegraphics[width=0.8\textwidth]{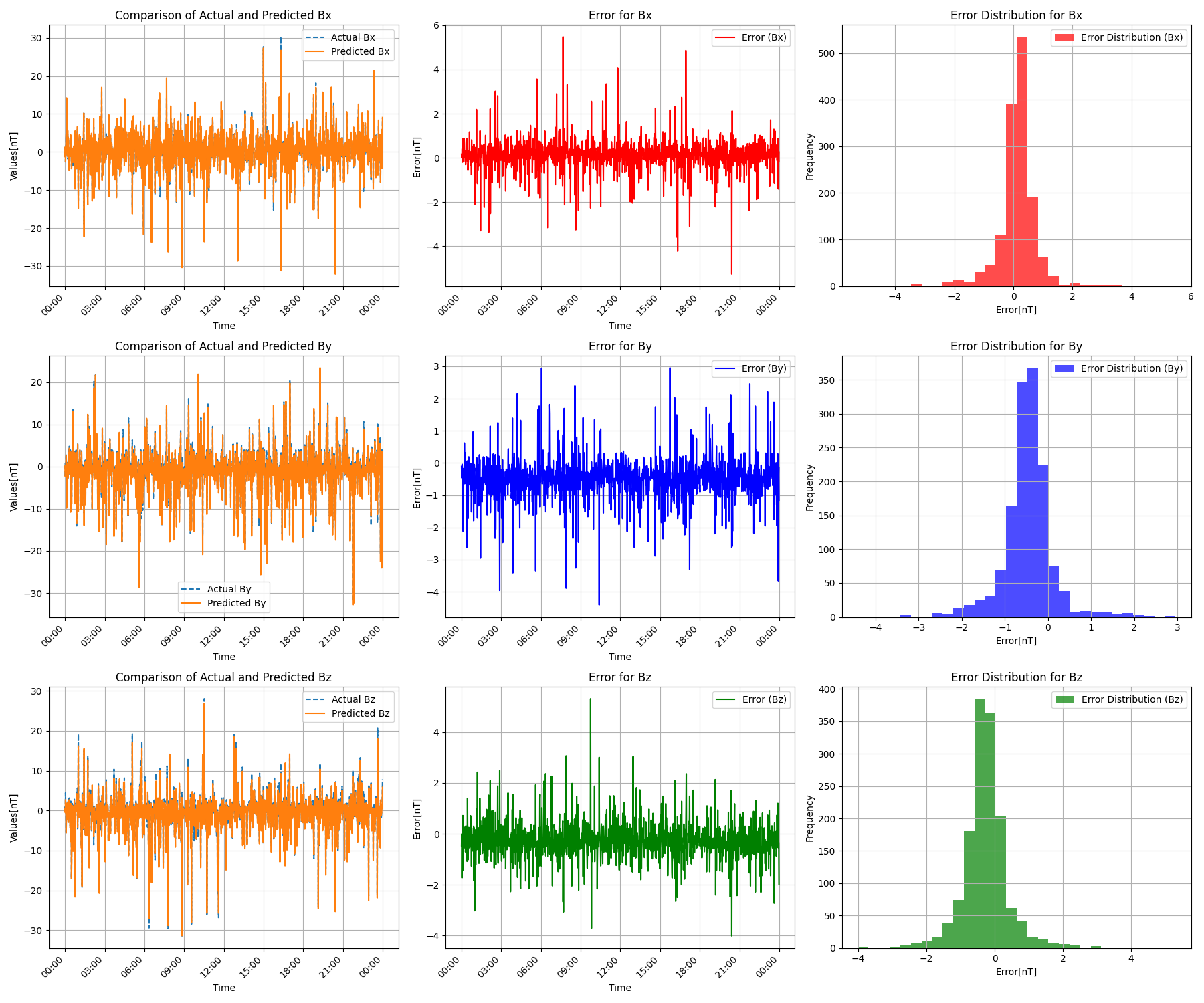}
  \caption{Comparison of Predicted Results and Actual Data on 2021-11-20}
  \label{fig:prediction_error_2021-11-20}
\end{figure}

\begin{figure}[H]
  \centering
  \includegraphics[width=0.8\textwidth]{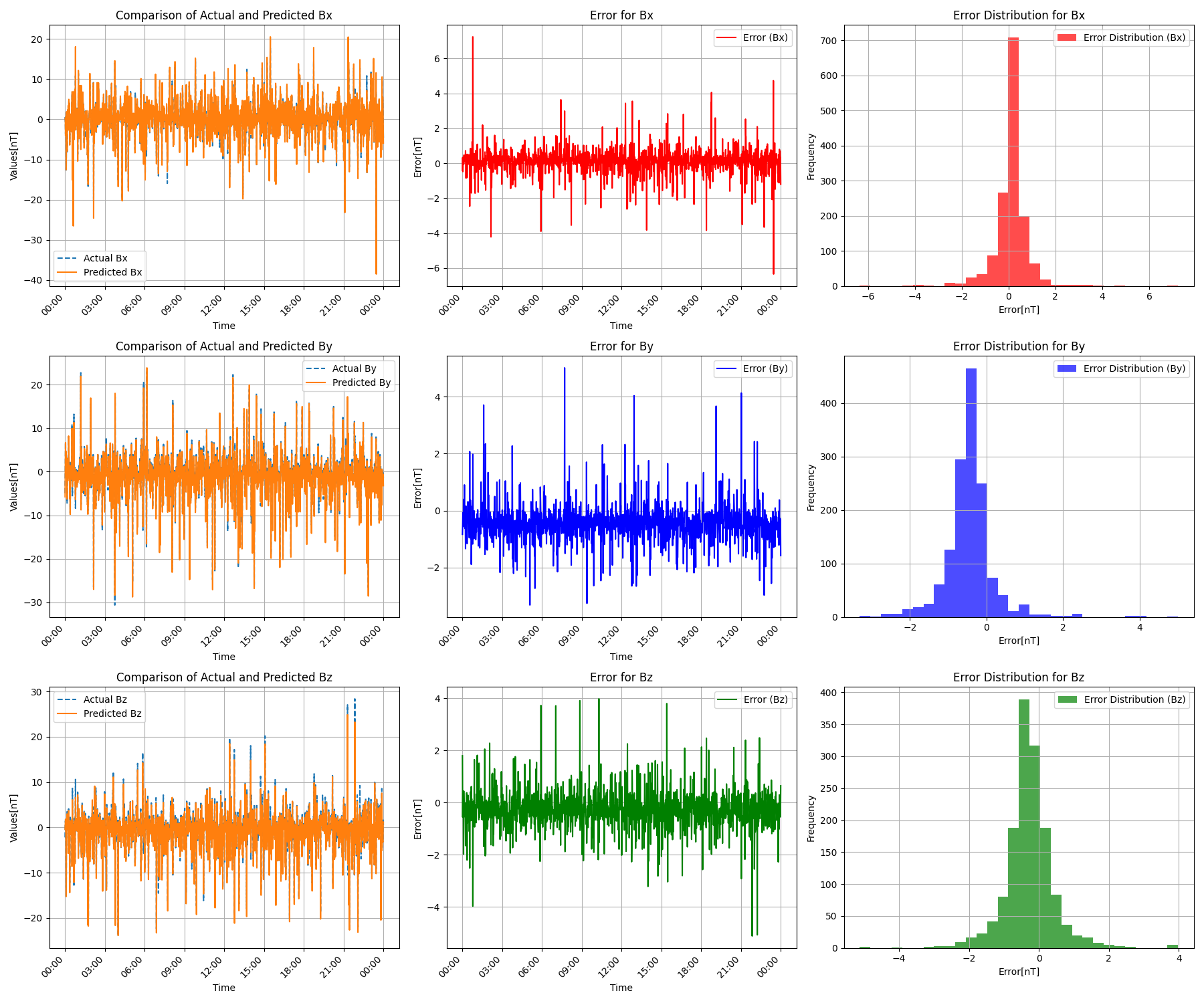}
  \caption{Comparison of Predicted Results and Actual Data on 2021-11-21}
  \label{fig:prediction_error_2021-11-21}
\end{figure}

\begin{figure}[H]
  \centering
  \includegraphics[width=0.8\textwidth]{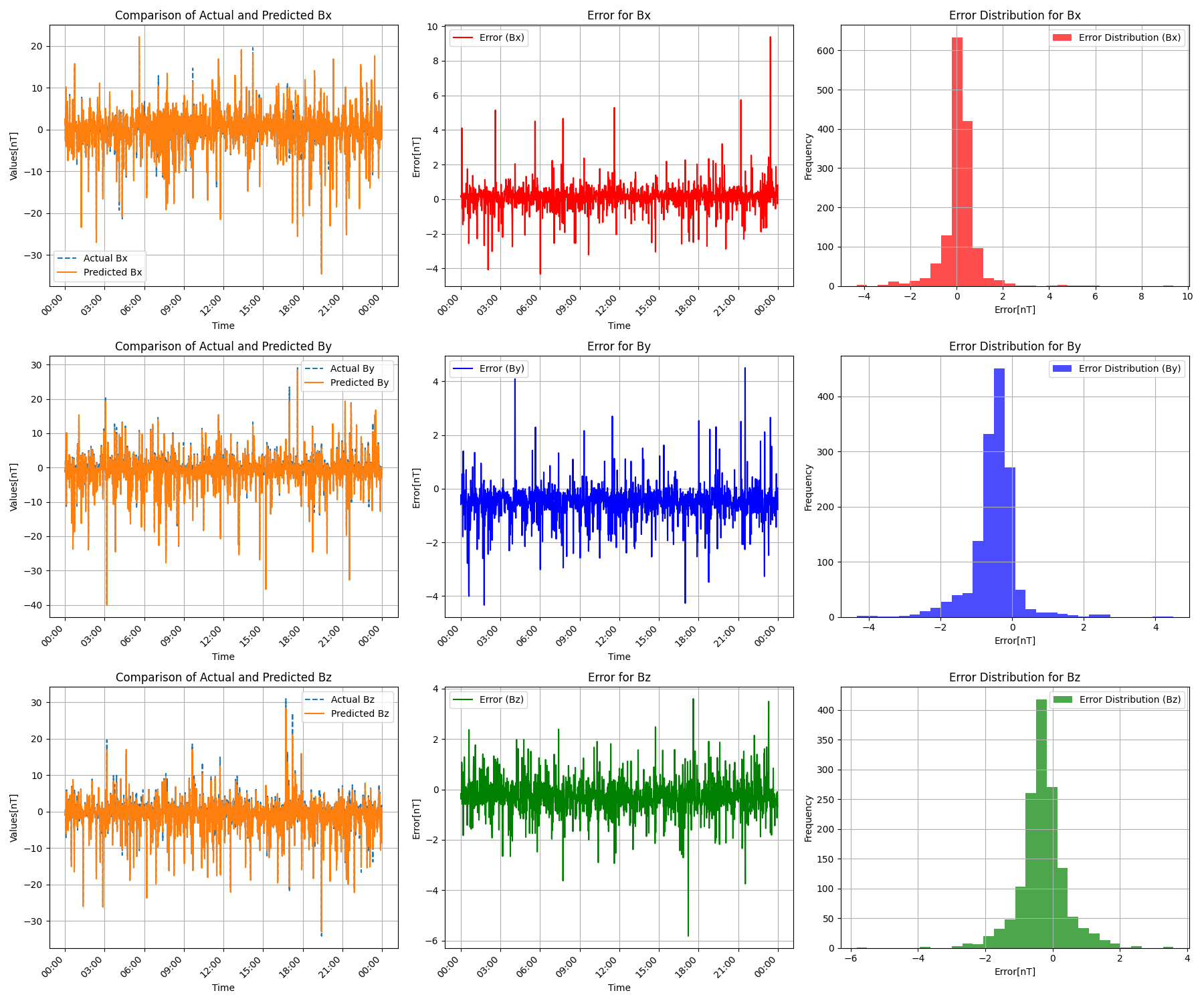}
  \caption{Comparison of Predicted Results and Actual Data on 2021-11-22}
  \label{fig:prediction_error_2021-11-22}
\end{figure}

\begin{figure}[H]
  \centering
  \includegraphics[width=0.8\textwidth]{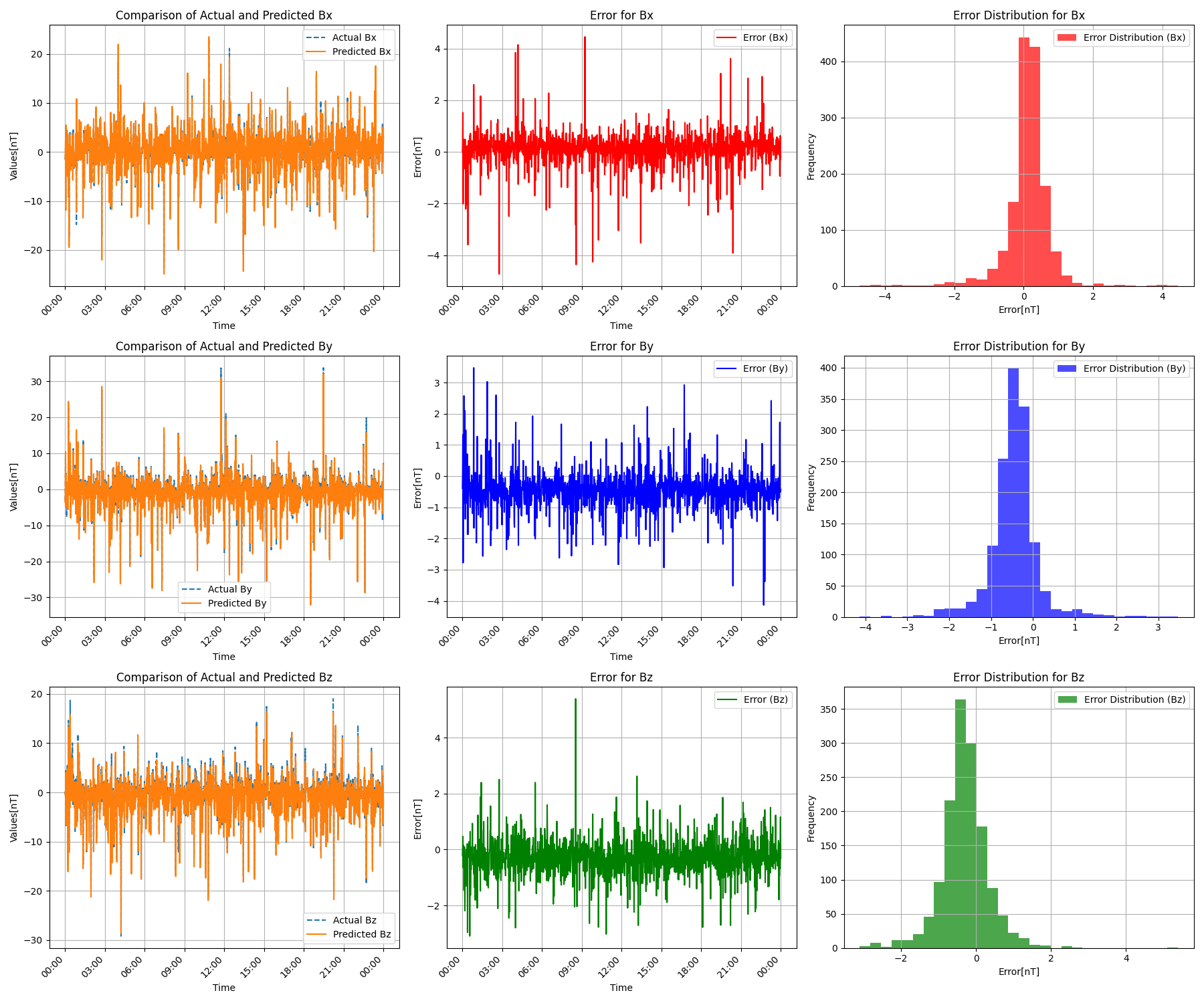}
  \caption{Comparison of Predicted Results and Actual Data on 2021-11-23}
  \label{fig:prediction_error_2021-11-23}
\end{figure}

\begin{figure}[H]
  \centering
  \includegraphics[width=0.8\textwidth]{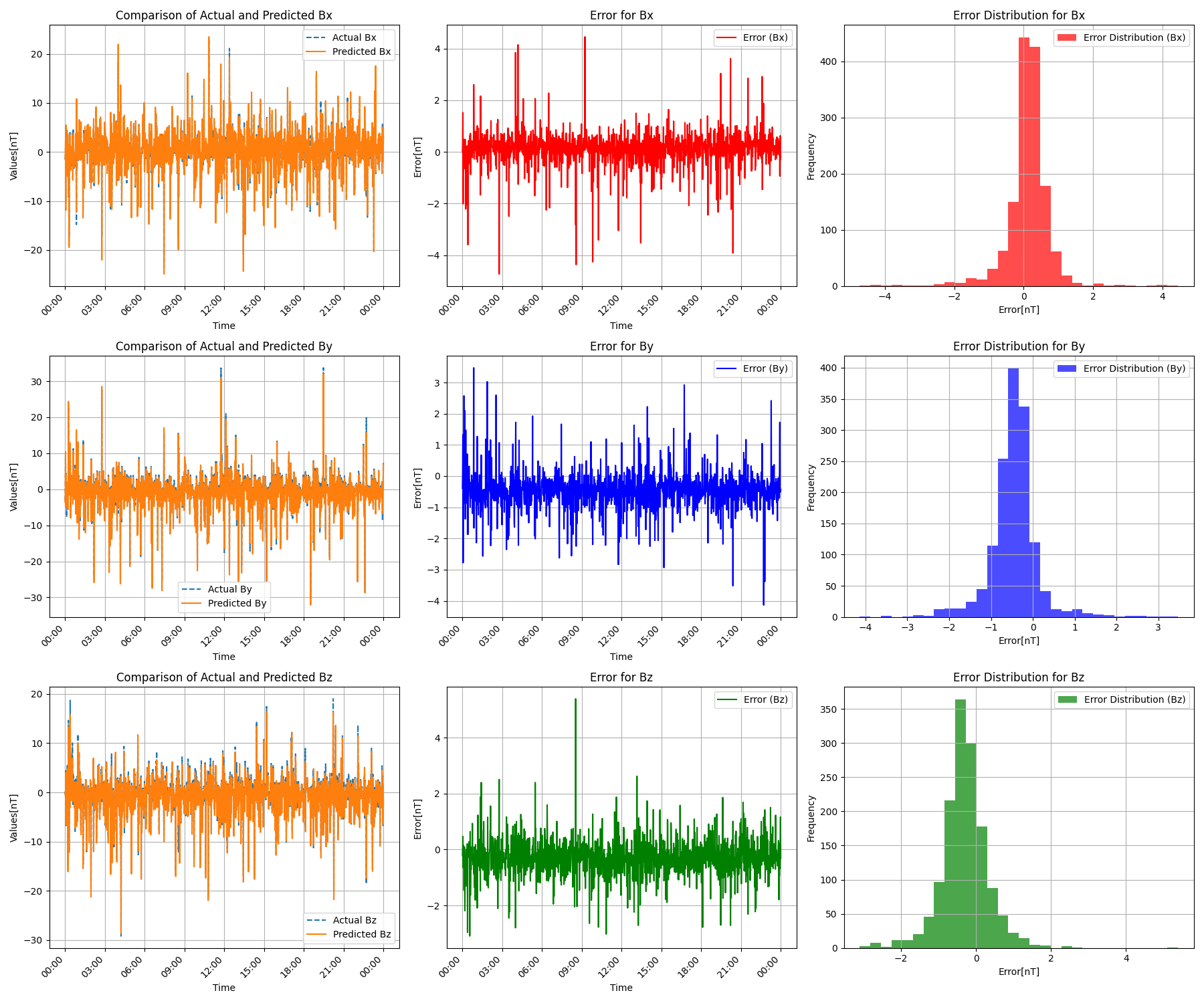}
  \caption{Comparison of Predicted Results and Actual Data on 2021-11-24}
  \label{fig:prediction_error_2021-11-24}
\end{figure}

We use two months data for training in 2021. We split the dataset $80\%$ as training set and $20\%$ as test set. Daily comparison plots of actual versus predicted values for \(B_x, B_y, B_z\) demonstrate how our methods effectively predict these components by directly comparing our predicted results with the actual data. Error distribution plots show reduced variance, indicating consistent model performance.
We have generated several plots to analyze the performance of our model by comparing our predicted results with the actual data using these methods. Figures~\ref{fig:prediction_error_2021-11-20}, \ref{fig:prediction_error_2021-11-21},  \ref{fig:prediction_error_2021-11-22}, \ref{fig:prediction_error_2021-11-23} and \ref{fig:prediction_error_2021-11-24} show the comparison of predicted results and actual data, along with the error analysis from our test set.

\begin{figure}[H]
  \centering
  \includegraphics[width=0.8\textwidth]{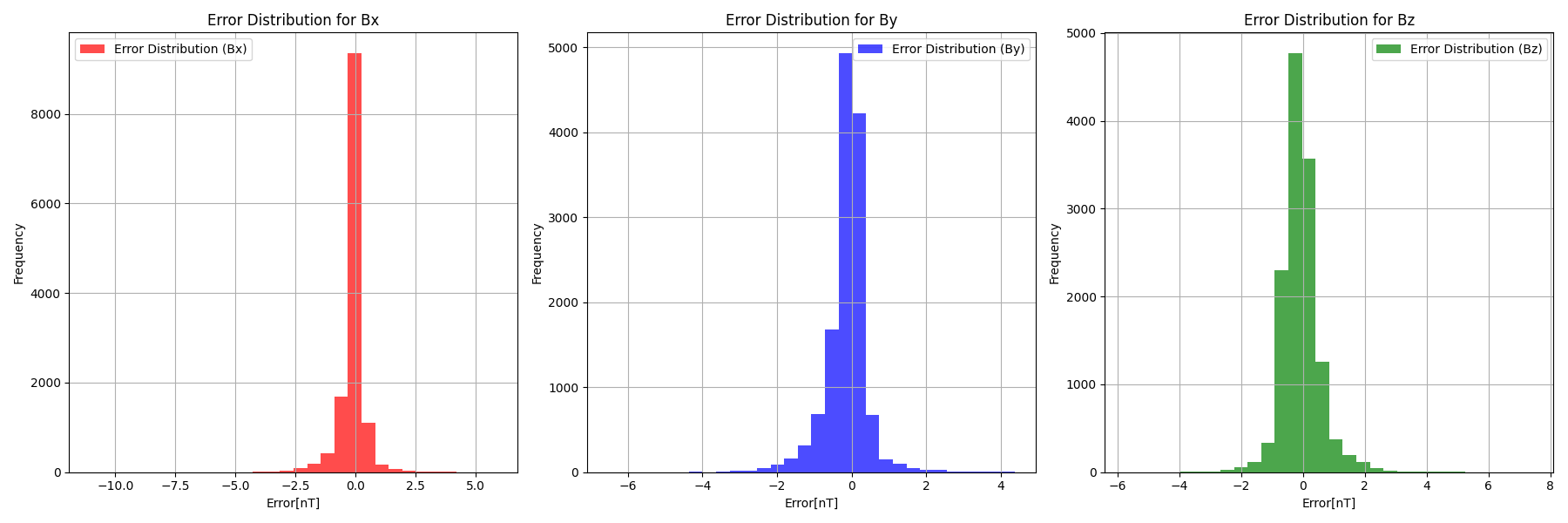}
  \caption{Transformer model error distribution on test set}
  \label{fig:Trans}
\end{figure}

\begin{figure}[H]
  \centering
  \includegraphics[width=0.8\textwidth]{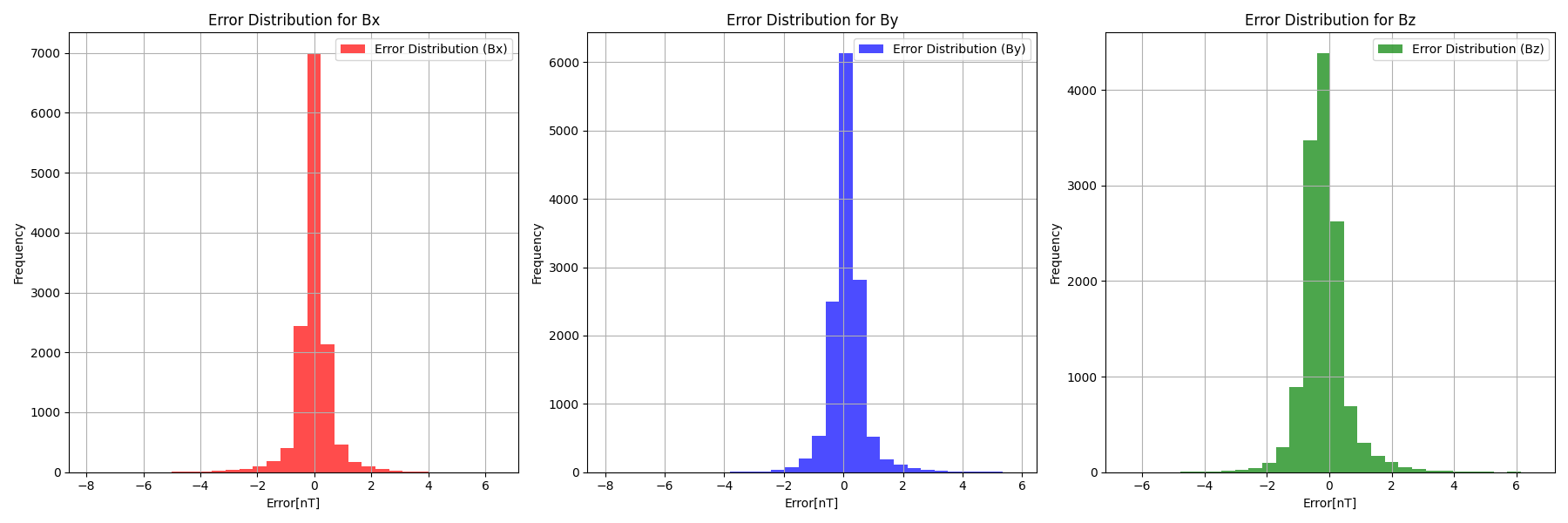}
  \caption{The Physics-Informed Transformer with Fourier Transform and Physics Constraint error distribution on test set}
  \label{fig:transtrans}
\end{figure}

The Physics Informed Transformer with Fourier Transform and Physics Constraint achieve a mean absolute error  0.44nT, and has similar results as the Transformer model. Here are the error distributions of the Transformer \ref{fig:Trans} and The Physics Informed Transformer \ref{fig:transtrans} on the test set.

\section{Discussion}
This calibration pipeline effectively reduces noise and anomalies in Tianwen-1 magnetometer data while drastically decreasing the data correction time from days or months to minutes or hours. 
\begin{itemize}
    \item \textbf{Significance of Magnetic Field Correction:} By systematically calibrating Tianwen-1 magnetometer data, our approach effectively resolves anomalies caused by satellite dynamics and instrument interference, substantially reducing the time required for correction.
    \item \textbf{Integration of Physics and Machine Learning:} Incorporating physical fields (electric and magnetic) as input features and employing a Transformer based model improves prediction accuracy and integrates fundamental physics directly into the machine learning framework.
    \item \textbf{Future Applications:} This modeling strategy for planetary magnetospheres can be extended to upcoming missions, such as Tianwen-2. It also enables improved forecasts of geomagnetic storms and solar wind interactions, contributing to broader space weather research.
    \item \textbf{Scalability and Real Time Prediction:} Our method is adaptable to other satellite missions and can be refined by incorporating additional physical parameters. Moreover, once trained, the model provides near-instantaneous predictions, delivering accurate, high quality calibrated data that can be readily integrated into mission support and decision making processes.    
\end{itemize}

\section{Conclusion}
This study introduces a scalable, physics informed machine learning framework that leverages Maxwell’s equations and a Transformer based neural network to rapidly and accurately correct magnetometer data from the Tianwen-1 Mars mission. While the original approach based on a standard Transformer already demonstrates significant improvements in both calibration speed and accuracy, our implementation of a second, physics informed Transformer further refines the predictions by integrating Fourier domain insights and enforcing a divergence free condition. By integrating fundamental physical constraints directly into the model, we improve the fidelity and consistency of the corrected magnetic field measurements. Compared to traditional calibration methods that rely on lengthy manual procedures and extended data segments, our approach requires only minutes to hours for training and provides near instantaneous predictions. The resulting data not only exhibit heightened accuracy and reduced noise but also facilitate more reliable analyses of planetary magnetospheric phenomena and solar wind interactions. Beyond Tianwen-1, this method serves as a flexible and robust blueprint for real time data calibration in future exploration missions, supporting advanced space weather modeling and significantly improving our ability to interpret complex electromagnetic environments in planetary systems.

\bibliographystyle{plain}
\bibliography{references.bib} 
\end{document}